\documentclass[aps,prd,twocolumn,superscriptaddress,notitlepage,nofootinbib]{revtex4-1}

\usepackage{amsmath}
\usepackage{graphicx}
\usepackage[normalem]{ulem} 
\usepackage{multirow} 
\usepackage{lipsum} 
\usepackage[usenames, dvipsnames]{xcolor} 
\usepackage[colorlinks=true,linkcolor=blue]{hyperref} 
\def\bea#1\eea{\begin{align}#1\end{align}} 

\newcommand{\bef}{\begin{figure}[htb]\centering}
\newcommand{\eef}{\end{figure}}
\allowdisplaybreaks

\newcommand{\jpsi}{{J/\psi}}

\newcommand{\state}[4]{{^{#1}\hspace{-0.6mm}{#2}_{#3}^{[#4]}}}

\newcommand\CScSa{\state{3}{S}{1}{1}}

\newcommand\COaSz{\state{1}{S}{0}{8}}

\newcommand\COcSa{\state{3}{S}{1}{8}}
\newcommand\COcPz{\state{3}{P}{0}{8}}

\newcommand\COcPj{\state{3}{P}{J}{8}}

\newcommand\mo{{\mathcal O}}

\newcommand{\LDME}[2]{\langle\mo^{#1}(#2)\rangle}

\newcommand{\mopa}{\LDME{\jpsi}{\COaSz}}

\newcommand\mopc{\LDME{\jpsi}{\COcPz}}
\newcommand\mopj{\LDME{\jpsi}{\COcPj}}

\newcommand{\LDMEn}[1]{\langle\mo^{#1}{(n)}\rangle}
\newcommand\mopn{\LDMEn{\jpsi}}

\begin{document}
\title{On the role of $\jpsi$ production in electron-ion collisions }

\date{\today}

\author{Zexuan Chu}
\affiliation{Key Laboratory of Nuclear Physics and Ion-beam Application (MOE) and Institute of Modern Physics, Fudan University,
Shanghai 200433, China}

\author{Jinhui Chen}
\email{chenjinhui@fudan.edu.cn}
\affiliation{Key Laboratory of Nuclear Physics and Ion-beam Application (MOE) and Institute of Modern Physics, Fudan University,
Shanghai 200433, China}
\affiliation{Shanghai Research Center for Theoretical Nuclear Physics, NSFC and Fudan University, Shanghai 200438, China}

\author{Xiang-Peng Wang}
\email{xiangpeng.wang@tum.de}
\affiliation{Technical University of Munich, TUM School of Natural Sciences, Physics Department, James-Franck-Strasse 1, 85748 Garching, Germany}

\author{Hongxi Xing}
\email{hxing@m.scnu.edu.cn}
\affiliation{Key Laboratory of Atomic and Subatomic Structure and Quantum Control (MOE), Guangdong Basic Research Center of Excellence for Structure and Fundamental Interactions of Matter, Institute of Quantum Matter, South China Normal University, Guangzhou 510006, China}
\affiliation{Guangdong-Hong Kong Joint Laboratory of Quantum Matter, Guangdong Provincial Key Laboratory of Nuclear Science, Southern Nuclear Science Computing Center, South China Normal University, Guangzhou 510006, China}
\affiliation{Southern Center for Nuclear-Science Theory (SCNT), Institute of Modern Physics, Chinese Academy of Sciences, Huizhou 516000, China}

\begin{abstract}
Within the framework of non-relativistic QCD (NRQCD) effective field theory, we study the leptoproduction of $J/{\psi}$ at next-to-leading order in perturbative QCD for both unpolarized and polarized electron-ion collisions. We demonstrate that the $J/{\psi}$-tagged deep inelastic scattering in the future Electron-Ion Collider can be served as a golden channel for the reasons including constraining NRQCD long distance matrix elements, probing the nuclear gluon distribution functions, as well as investigating the gluon helicity distribution inside a longitudinal polarized proton.
\end{abstract}

\maketitle

\section{Introduction}
$\jpsi$ production has been the focus of much theoretical and experimental interest in the past decades. Understanding its production mechanism and using it as a probe to QCD matter are among the most active and challenging subjects in particle and nuclear physics \cite{Rothkopf:2019ipj,Chapon:2020heu}. Among several extensively used models and effective field theories for the description of heavy quarknoium production, the most sophisticated one in high energy collisions is provided by non-relativistic QCD (NRQCD) effective field theory \cite{Bodwin:1994jh}, in which the $\jpsi$ hadronization mechanism is encoded in the non-perturbative long distance matrix elements (LDMEs). In the past decades, through explicit perturbative calculation of $\jpsi$ production at next-to-leading order (NLO) in electron-positron, proton-proton and electron-proton collisions within the NRQCD factorization framework, significant progress has been made to extract the LDMEs from world data, see e.g. \cite{Bodwin:2014gia,Butenschoen:2011yh,Chao:2012iv,Gong:2012ug,Bodwin:2015iua,Brambilla:2022rjd,Brambilla:2022ayc}. However, no consensus has been reached yet to identify the universalities of these LDMEs and to understand the polarization of the produced $\jpsi$ \cite{Lansberg:2019adr,Chen:2021tmf}. 

On the other hand, taking advantages of the fundamental properties of heavy quarkonium and its production mechanism, it has been proposed that heavy quarkonium can be served as a sensitive probe of the fundamental properties of nuclear matter. Remarkable examples including the identification of quark gluon plasma \cite{Matsui:1986dk, Tang:2020ame,Wang:2022fwq} and probing the gluon jet quenching effect in heavy ion collisions \cite{Zhang:2022rby,Zhang:2024owr}, searching for signatures of gluon saturation in ultraperipheral collisions \cite{CMS:2023snh}, as well as the study of short range correlation in nucleus \cite{Xu:2019wso}. Moreover, with the advent of electron-ion colliders \cite{Accardi:2012qut,Anderle:2021wcy}, new proposals of using $\jpsi$ production in lepton-proton collisions to probe the nucleon structure have been put forward recently, see e.g. \cite{DAlesio:2021yws,Boer:2021ehu,Qiu:2020xum,Chen:2023hvu,Flore:2020jau,Bacchetta:2018ivt,Copeland:2023qed,liu:2023unl}. 
All these proposals are driven by the prominent feature that $\jpsi$ in deep inelastic scattering (DIS) is dominated by virtual photon-gluon fusion process, which can be inferred from the leading order diagram as shown in Fig. \ref{fig-diagram}(a). Such a unique feature drives the expectation that $\jpsi$ production is more sensitive to the gluon content of nucleon and nucleus, comparing to the generally recognized observables like dijet production \cite{Boer:2016fqd,Xing:2012ii}, which suffers contamination from virtual photon-quark channel. 

Similar feature also exists in $\jpsi$ photoproduction that has been measured by HERA \cite{ZEUS:2002src,H1:2010udv,H1:2002voc} and complete NLO calculations are performed in Refs. \cite{Butenschoen:2009zy,Artoisenet:2009xh,Chang:2009uj}. In this paper, we emphasize the exceptional role of $\jpsi$-tagged DIS in future Electron-Ion Collider (EIC) \cite{Accardi:2012qut,AbdulKhalek:2021gbh}. In particular, through explicit calculation at NLO for $\jpsi$ production in unpolarized electron-proton ($ep$) collisions, we explore the novel opportunities to constrain the LDMEs $\mopa$ and $\mopc$ considering the high luminosity of EIC, which could provide new insights into the $\jpsi$ production mechanism. In addition, we present numerical result for the nuclear modification factor at NLO for electron-nucleus ($eA$) collisions, we find out that the $\jpsi$-tagged DIS could be an excellent probe to gluon collinear distribution of nucleus. Furthermore, we extend such a calculation to polarized $ep$ collision at NLO for the first time. Numerical results on double longitudinal spin asymmetry are presented to test the sensitivity to nucleon helicity distributions. The results presented in this paper strongly motivate the measurement of $\jpsi$-tagged DIS at the future EIC.

\section{NLO calculation for $\jpsi$ production at EIC}
We start with prompt $\jpsi$ production in $ep$ collision, 
\bea
e(\ell)+P(p) \to e(\ell^{\prime})+\jpsi + X,
\eea
where $\ell, p$ and $\ell^{\prime}$ are, respectively, the momentum of the incoming electron, proton and the outgoing electron. In particular, we focus on the $\jpsi$-tagged DIS event, where the phase space for the prompt $\jpsi$ is integrated out.
Within the framework of NRQCD and leading twist collinear factorization, the  differential cross section for the $\jpsi$-tagged DIS can be written in the following form \cite{Fleming:1997fq}
\bea
\label{eq:NRQCDfac}
\frac{d\sigma}{dQ^2 dy} =& \sum_i \sum_{n} \int d x f_{i/p}(x,\mu_F^2)  C_i(n)\mopn,
\eea
where $Q^2=-q^2$ with $q=\ell-\ell^{\prime}$ stands for the momentum of the virtual photon, $y=p\cdot q/p\cdot \ell$, $f_{i/p}$ is the proton parton distribution functions (PDFs), $x$ is the momentum fraction of proton carried by parton $i$ and $\mu_F$ is the factorization scale. $C_i(n)$ is the short-distance coefficient with initial parton $i$ and final Fock state $c\bar c[n]$, $\mopn$ is the LDME representing the hadronization of $c\bar c[n]$ Fock state to the final observed $\jpsi$, which has definite power counting of $v$, the relative velocity of the heavy quark and anti-quark in the quarkonium rest frame. Up to $\mathcal{O}(v^4)$ in $v$ expansion, without considering the $v^2 $ suppressed relativistic corrections for the color-singlet channel, the $\CScSa, \COaSz, \COcSa, \COcPj$ channels will contribute to the factorization formula in Eq. (\ref{eq:NRQCDfac}). For the $\COcPj$ channel, we use the relation $\mopj =(2J+1)\mopc$ to convert all the P-wave LDMEs to $\mopc$, so that $C_i(\COcPj)$ represents the   P-wave short-distance coefficients summed over $J =0, 1, 2$.

\begin{figure}[htbp]
	\centering	\includegraphics[width=0.5\textwidth]{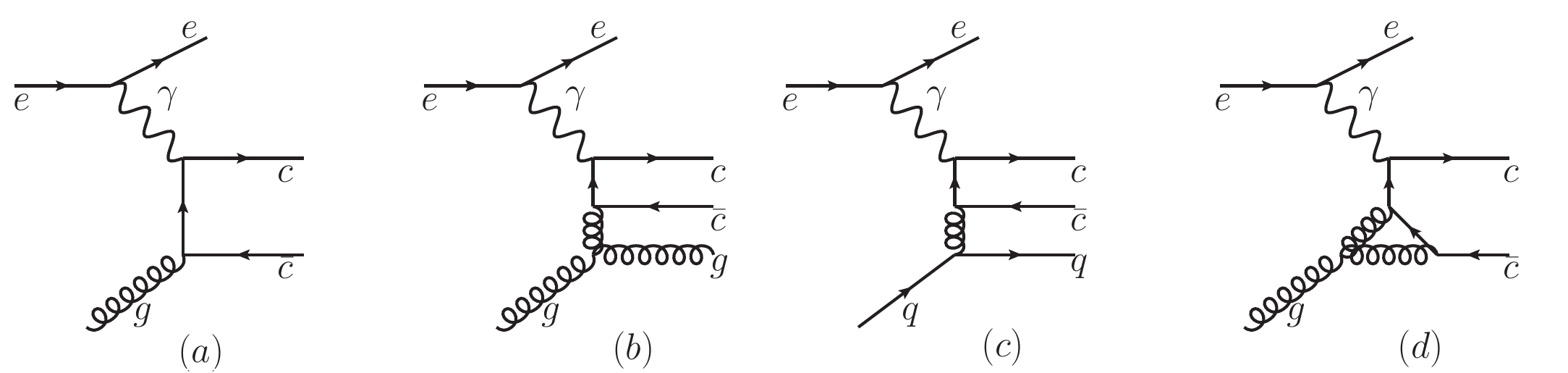}
\caption{Representative Feynman diagrams for LO (a), real correction from gluon (b) and quark (c) channel at NLO, and virtual correction (d) at NLO for $\jpsi$ production in DIS. 
}
\label{fig-diagram}
\end{figure}

As shown in Fig. \ref{fig-diagram}(a), the $\jpsi$-tagged DIS at LO is solely determined by gluon channel. Therefore, only the color octet intermediate states $c\bar{c}(\COaSz)$ and $c\bar{c}(\COcPj)$ contribute and the corresponding short distance coefficients can be expressed as \cite{Fleming:1997fq}
\bea
C_g^{(0)}(\COaSz )=&\frac{1+(1-y)^2}{y}C_0, \\
C_g^{(0)}(\COcPj)=&\Bigg[\frac{1+(1-y)^2}{y}\frac{3Q^2/m^2+28}{Q^2+4m^2}\nonumber\\
&\,\, \,  -\frac{32yQ^2}{(Q^2+4m^2)^2}\Bigg] C_0,
\eea 
where $C_0 = \frac{2\pi^2\alpha_s\alpha^2e_c^2}{Q^2(Q^2+4m^2)m}\delta(xyS-Q^2-4m^2)$ with $m$ the charm quark mass and $S=(\ell+p)^2$, $e_c$ is the charm quark electrical charge, $\alpha_s$ and $\alpha$ are, respectively, the QCD and QED coupling constant. 

According to the LO result, one would expect that the $\jpsi$-tagged DIS can be uniquely used to determine the gluon distribution in proton. However, such a unique feature could be contaminated by the quark channel, which appears at higher orders. In order to test such an effect and check the convergence of higher order corrections, we perform the first complete NLO calculation for $\jpsi$-tagged DIS, that is at $\mathcal{O}(\alpha^2\alpha_s^2)$. At NLO, in addition to the real and virtual corrections for $\COaSz$ and $\COcPj$ channels from quark and gluon channels as shown in the representative diagrams in Fig. \ref{fig-diagram} (b-d), one also needs to include contributions from both $^3S_1^{[1]}$ and $^3S_1^{[8]}$ channels, where the former only exists in gluon channel. In our calculation, the Feynman amplitudes are generated using FeynArts \cite{Hahn:2000kx}, the color and Dirac algebra are performed with the help of FeynCalc \cite{Mertig:1990an}, which results in 4 diagrams for  quark channel and 8 diagrams for gluon channel in real correction, and 17 diagrams for virtual correction including 6 counter term diagrams. 

We use dimensional regularization ($D=4-2\epsilon$) to regularize infrared and collinear divergences in real correction, and infrared, collinear and ultraviolet (UV) ones in virtual correction. In particular, HVBM scheme \cite{tHooft:1972tcz} is implemented to deal with $\gamma_5$. In real correction, we integrate over the whole phase space of the outgoing particles except for the outgoing electron, which results in soft ($\sim 1/\epsilon$), collinear ($\sim 1/\epsilon$) and soft-collinear ($\sim 1/\epsilon^2$) divergences. Regarding virtual corrections, we use the packages Apart \cite{Feng:2012iq} and FIRE \cite{Smirnov:2014hma} for the reduction of loop integrals. The derived UV divergences are removed through renormalization procedure. Notice that we adopt a mixed renormalization scheme\cite{Klasen:2004az}, in which the renormalization constants of the heavy quark mass ($Z_m$), heavy quark field ($Z_2$) and gluon field ($Z_3$) are defined in the on-shell scheme, while the renormalization constant of the strong coupling ($Z_g$) is defined in $\rm \overline {MS}$ scheme, thus the self-energy diagrams for the external fields are excluded in the computation. The soft, collinear and soft-collinear divergences are cancelled in the sum of real and virtual corrections or absorbed into the redefinition of the nucleon/nucleus PDF $f_{g/p}(x,\mu_F^2)$.
Since we focus on the DIS event with $Q^2>0$ in our study, there will be no divergence resulting from the QED splitting $e\rightarrow e\gamma$ \cite{Qiu:2020xum}.  

Likewise, one can also calculate the cross section for $\jpsi$-tagged DIS at NLO in longitudinal polarized collisions. As we expected, one encounters the same singularities as in unpolarized case, except for the redefinition of longitudinal polarized PDFs $\Delta f_{i/p}(x,\mu_F^2)$. After the regularization of all singularities, we are left with the finite part, which results into the following differential cross section at NLO,
\bea
\frac{d\Delta\sigma}{dQ^2 dy} =& \sum_i \sum_{n} \int d x \Delta f_{i/p}(x,\mu_F^2)  \Delta C_i(n)\mopn.
\eea

 In this calculation, we obtain for the first time the fully analytical expressions for the short distance coefficients $C_i(n)$ for unpolarized and $\Delta C_i(n)$ for longitudinal polarized cases at NLO \footnote{Refer to an ancillary file attached to the arXiv submission for the full analytical expressions of $C_i(n)$ and $\Delta C_i(n)$.}. Such analytical results will be very efficient in the global analysis of LDMEs and PDFs for nucleon/nucleus when experimental data are available.

\section{$\jpsi$ as a probe to nonperturbative functions}
In this section, we show phenomenological results by employing the obtained NLO analytical expressions under future EIC designs. We choose a center-of-mass energy of $\sqrt{s} = 100$ GeV to illustrate the power of $\jpsi$ as a probe to nonperturbative functions. In our numerical calculation, the heavy quark mass $m=1.55$ GeV and the number of light quark flavor $n_f=3$, the factorization scale $\mu_F$ and the renormalization scale $\mu_R$ are chosen as $\mu_F=\mu_R=\sqrt{Q^2+4m^2}$. 

\subsection{Sensitivity to NRQCD LDMEs}
We first present the dependence of $\jpsi$-tagged DIS cross section on the NRQCD LDMEs, which are determined through global analysis of world data at NLO accuracy within the framework of NRQCD. With different fitting strategies, various sets of $\jpsi$ LDMEs are obtained, challenges the universality of these nonperturbative functions. In this work, we adopt the LDMEs from five groups: Bodwin {\it et al.} \cite{Bodwin:2014gia}, Butenschoen {\it et al.} \cite{Butenschoen:2011yh}, Chao {\it et al.} \cite{Chao:2012iv}, Gong {\it et al.} \cite{Gong:2012ug} and Brambilla {\it et al} \cite{Brambilla:2022rjd,Brambilla:2022ayc}. The optimized values are summarized in Table \ref{table:nrqcd}. 

   \begin{footnotesize}
\begin{table}[hbt]
\caption{$\jpsi$ NRQCD LDMEs from five different groups.}
\label{table:nrqcd}
\begin{center}
  \begin{tabular}{  l |c | c | c | c  }
  \hline
    & $\langle {\mathcal O}(^3S_1^{[1]})\rangle$ & $\langle {\mathcal O}(^1S_0^{[8]})\rangle$ & $\langle {\mathcal O}(^3S_1^{[8]})\rangle$ & $\langle {\mathcal O}(^3P_0^{[8]})\rangle$ \\ 
    & GeV$^3$ & $10^{-2}$ GeV$^3$ & $10^{-2}$ GeV$^3$ & $10^{-2}$ GeV$^5$ 
        \\ \hline
    Bodwin & 0 & 9.9 & 1.1  & 1.1 \\ \hline
       Butenschoen & 1.32 & 3.04 & 0.16 & $-0.91$\\ \hline
   Chao & 1.16 & 8.9 & 0.30  & 1.26 \\ \hline
   Gong & 1.16 & 9.7 & $-0.46$ & $-2.14$ \\ \hline
   Brambilla & 1.18 &  $-0.63$ & 1.4 & 5.83 \\ \hline
  \end{tabular}
\end{center}
\end{table}
  \end{footnotesize}
  
The LDMEs from Bodwin {\it et al.}, Chao {\it et al.} and Gong {\it et al.} give $^1S_0^{[8]}$ dominance scenario and thus naturally conform the almost unpolarized $J/\psi$ hadron production data at large $p_T$. However, the $^1S_0^{[8]}$ dominance scenario contradicts with the $\eta_c$ hadron production data under the assumptions of heavy quark spin symmetry (HQSS)\cite{Butenschoen:2014dra,Han:2014jya,Zhang:2014ybe,Bodwin:1994jh}. 
The LDMEs from Butenschoen {\it et.al.} lead to the $J/\psi$ polarization puzzle problem \cite{Brambilla:2010cs}. Although, the LDMEs from Brambilla {\it et.al} do not have the above issues, $\mopa$ from Brambilla {\it et.al} is poorly constrained and suffers from large uncertainties.

\begin{figure}[htbp]
	\centering	\includegraphics[width=0.45\textwidth]{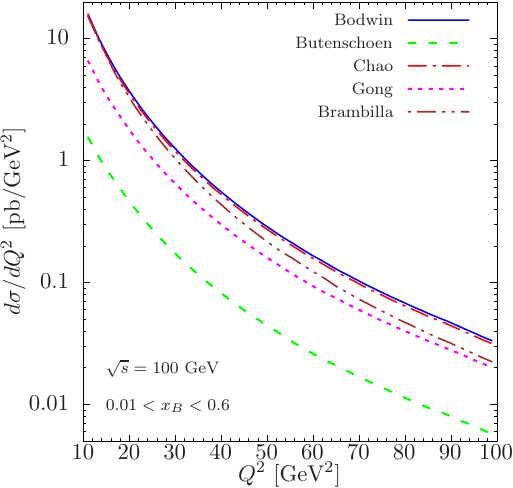}
\caption{NLO differential cross section as a function of $Q^2$ for $\jpsi$-tagged DIS process.}
\label{fig-xsec}
\end{figure}

In Fig.\ref{fig-xsec}, we plot the $\jpsi$-tagged DIS differential cross section as a function of $Q^2$ using CT18NLO \cite{Hou:2019efy} for proton PDFs and the five sets of LDMEs shown in Table \ref{table:nrqcd}. We find that the LDMEs from five groups lead to tremendous differences. We have checked that the differential cross sections are dominated by $\COaSz$ and $\COcPj$ channels.
The values of the LDMEs $\langle {\mathcal O}^{J/\psi}(^1S_0^{[8]})\rangle, \langle {\mathcal O}^{J/\psi}(^3P_0^{[8]})\rangle$  from the groups of Bodwin and Chao are very close, thus leads to almost identical predictions, which are an order of magnitude larger compared with the predictions obtained using the LDMEs from Butenschoen {\it et al.} and significantly larger than those derived using the LDMEs from Gong {\it et al.}. The LDMEs from Brambilla {\it et al.} give similar predictions with Bodwin {\it et al.} and Chao {\it et al.} at low $Q^2$ but gradually converge towards those using the LDMEs from Gong {\it et al.} at large $Q^2$, which leads to a distinct shape of $Q^2$ dependence. The drastic difference of $\jpsi$-tagged DIS differential cross section evaluated with the LDMEs from five different groups, clearly demonstrates that this observable is a very sensitive probe of NRQCD LDMEs, especially for $\langle {\mathcal O}^{J/\psi}(^1S_0^{[8]})\rangle$ and $\langle {\mathcal O}^{J/\psi}(^3P_0^{[8]})\rangle$.

\subsection{Sensitivity to nuclear PDFs}
\begin{figure}[htbp]
	\centering	\includegraphics[width=0.45\textwidth]{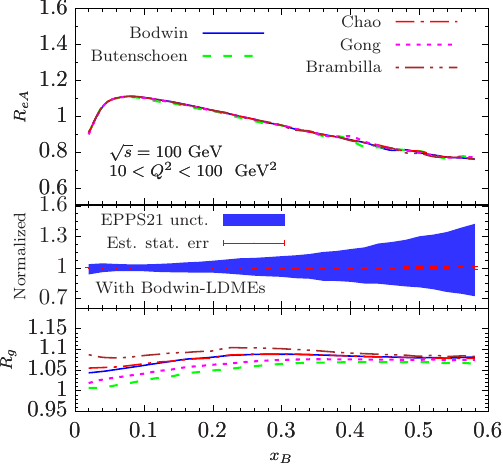}
\caption{The nuclear modification factor as a function of $x_B$ for leptoproduction of $\jpsi$ in $ePb$ collision at $\sqrt{s} = 100$ GeV.}
\label{fig-ReA}
\end{figure}

The high luminosity of electron-nucleus program at the EIC, combined with the gluon-dominated $\jpsi$-tagged DIS process, promises profound insights into nuclear gluon distribution. To delve deeper into the role of the $\jpsi$-tagged DIS process at the EIC in elucidating nuclear partonic structure, we investigate the nuclear modification factor defined as follows,
\bea
R_{eA} = \left. \frac{d\sigma^{eA}}{dx_B} \right/ \frac{d\sigma^{ ep}}{dx_B},
\eea
in which we have adopted EPPS21 parametrization \cite{Eskola:2021nhw} for nuclear PDFs and CT18NLO for proton PDFs. In the top panel of Fig. \ref{fig-ReA}, we present the nuclear modification factor plotted against the Bjorken variable $x_B=Q^2/p\cdot q$. We observe that the five sets of LDMEs yield nearly identical results for $R_{eA}$, indicating the insensitivity of LDME choices and substantially reducing theoretical uncertainty arising from LDME variations.
In the middle panel of Fig. \ref{fig-ReA}, we use the LDMEs from Bodwin {\it et.al.} as an example to show the normalized (by central values) uncertainty of the predicted $R_{eA}$. The blue band represents normalized uncertainty resulting from the nuclear PDFs uncertainty using EPPS21 parametrization, from which we can see that the relative uncertainty increases as $x_B$ getting larger and becomes significant when $x_B$ approaching $0.6$. The normalized uncertainty coming from the statistical error, which is based on the best fit of EPPS21 and the expected EIC integrated luminosity 10 $fb^{-1}/A$ for $eA$ collisions~\cite{AbdulKhalek:2021gbh} and the detector response simulation~\cite{Li:2022kwn}, is shown in the vertical red bars. It is obvious that the uncertainty coming from the estimated statistical error is negligible comparing to those resulting from the EPPS21 uncertainty. It is important to note that the $eA$ differential cross section is dominated by the gluon channel in all region of $x_B$ considered as it was shown in the bottom panel of Fig. \ref{fig-ReA}, where $R_g$ is defined as the ratio of $\jpsi$ tagged differential cross sections in $eA$ collision without and with contributions from quark channels. The fact that the gluon channel gives $100\%$ to $110\%$ of the differential cross section calculated at $\mathcal{O}(\alpha^2\alpha_s^2)$, which means the quark channels only give small negative contribution at $\mathcal{O}(\alpha^2\alpha_s^2)$, strongly indicates that the $\jpsi$-tagged DIS process can be served as a clean probe to access the nuclear gluon distribution at the future EIC. 

\subsection{Sensitivity to nucleon helicity distribution}

We extend our NLO calculation of leptoproduction of $\jpsi$ from unpolarized to polarized $ep$ collisions, aiming at probing the proton gluon helicity distribution, which is of significant importance in understanding the configuration of proton spin \cite{Ji:2020ena}. 
\begin{figure}[htbp]
	\centering	\includegraphics[width=0.45\textwidth]{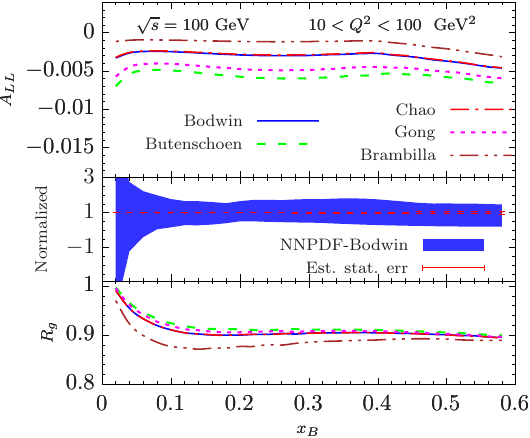}
\caption{The double longitudinal spin asymmetry as a function of $x_B$ for leptoproduction of $\jpsi$ in $ep$ collisions at $\sqrt{s} = 100$ GeV.}
\label{fig-ALL}
\end{figure}

We focus on the double longitudinal spin asymmetry for leptoproduction of $\jpsi$ defined as
\bea
A_{LL}  = \frac{ d\sigma(+,+) - d\sigma(+,-) }{d\sigma(+,+) + d\sigma(+,-)}=\frac{d\Delta\sigma}{d\sigma},
\eea
where the first and the second signs ($+, -$) in the parentheses right after $d\sigma$ represent the helicity of the incoming lepton and proton, respectively. 
We apply NNPDF parametrization \cite{Nocera:2014gqa}  for the longitudinal polarized proton PDFs.  

As we can see from the top panel of Fig. \ref{fig-ALL}, the predicted central value of $A_{LL}$ with the LDMEs sets from five groups are considerably different, except for those from Bodwin {\it et al.} and Chao {\it et al.} due to their close values of $\langle {\mathcal O}^{J/\psi}(^1S_0^{[8]})\rangle, \langle {\mathcal O}^{J/\psi}(^3P_0^{[8]})\rangle$. Once the values of $J/\psi$ LDMEs are fixed, one can use the theory predictions of $A_{LL}$ to constrain the gluon helicity distribution since $A_{LL}$ is dominated by gluon channel, as it was shown in the bottom panel of Fig. \ref{fig-ALL}, where $R_g$ is defined as the ratio of $d\Delta\sigma$ evaluated without and with the contributions from quark channels, respectively. 
The middle panel of Fig.~\ref{fig-ALL} shows the relative large uncertainty coming from lacking of precise knowledge on the gluon helicity distribution. This can be reduced significantly when we compare the $A_{LL}$ data in the future EIC with integrated luminosity 100 $fb^{-1}$ for $ep$ collisions~\cite{AbdulKhalek:2021gbh} with theory predictions.  

\section{Summary}
We calculate the leptoproduction of $\jpsi$ with five sets of LDMEs at NLO, under both unpolarized and polarized situations. The tremendous difference arises from various LDMEs indicates that the $\jpsi$-tagged DIS is a very sensitive probe to solve the problem of the universality of NRQCD LDMEs by comparing with the future EIC data. Besides, considering both the high collision luminosity and highly polarized beams in future EIC, as well as the dominance of gluon channel in $\jpsi$-tagged DIS process, this process can be also served as an unique opportunity to probe the nuclear gluon distribution and polarized gluon distribution. All these aspects strongly motivate the measurement of $\jpsi$-tagged DIS at the future EIC.

{\it Acknowledgement.}
This work was supported in part by the Guangdong Major Project of Basic and Applied Basic Research No. 2020B0301030008 and 2022A1515010683, by the National Natural Science Foundation of China (NSFC) (Nos. 12025501, 12035007 and 12022512). The work of  X.-P. W. is supported by the
DFG (Deutsche Forschungsgemeinschaft, German Research Foundation) Grant No. BR4058/2-2. X.-P. W. acknowledges support from the DFG cluster of excellence “ORIGINS” under
Germany’s Excellence Strategy - EXC-2094 - 390783311. X.-P. W. acknowledges support from STRONG-2020- European Union’s Horizon 2020 research and innovation program
under grant agreement No. 824093.


\end{document}